\documentclass[12pt,a4paper,twoside,english]{article}
     \usepackage[latin1]{inputenc}
     \usepackage[T1]{fontenc}
     \usepackage[usenames,dvipsnames]{color}
     \usepackage{graphicx}
     \usepackage{babel}
     \usepackage{verbatim}
     \usepackage{amsmath}
     \usepackage[nomarkers]{endfloat}
     \usepackage{epsf,rotate,a4wide}
\newcommand{\lp}{\left(}
\newcommand{\rp}{\right)}
\newcommand{\lc}{\left[}
\newcommand{\rc}{\right]}
\newcommand{\la}{\left \{ }
\newcommand{\ra}{\right \} }
\newcommand{\dr}{\partial}
\newcommand{\pca}{{\cal P}}
\newcommand{\hpca}{\hat{\cal P}}
\newcommand{\hd}{\hat{d}}

\newcommand{\drti}{\tilde{\dr}}
\newcommand{\xxp}{\Lambda_+}
\newcommand{\xxm}{\Lambda_-}
\newcommand{\xxa}{\widetilde{\Lambda}}
\newcommand{\xxb}{\widetilde{\widetilde{\Lambda}}}

\newcommand{\clst}{{\cal L}^*}
\newcommand{\clba}{\overline{\cal L}}

\newcommand{\cxst}{{\cal X}^*}
\newcommand{\cxba}{\overline{\cal X}}

\newcommand{\nd}{{\sf d}}

\newcommand{\bnabla}{\mbox{\boldmath $\nabla$}}

\newcommand{\Real}{{\rm I \! \! \! \; R}}

\addtocontents{lof}{\protect\vspace{0.5cm}}

\newcommand{\cxha}{\widehat{\cal X}}
\newcommand{\tba}{\overline{T}}
\newcommand{\pisb}{\overline{\pi}_s}

\newcommand{\nba}{\overline{n}}
\newcommand{\hba}{\overline{H}}
\newcommand{\muba}{\overline{\mu}}
\newcommand{\must}{\mu^*}
\newcommand{\cz}{{\cal Z}}

\newcommand{\qba}{\overline{q}}

\newcommand{\Ht}{{H_D}}

\begin{document}

\newlength{\figwidth}
\setlength{\figwidth}{0.9\textwidth}

%

\vspace*{0.5cm}
\normalsize

\centerline{\bf Stability of Leap-Frog Constant-Coefficients Semi-Implicit Schemes for}
\medskip
\centerline{\bf the Fully Elastic System of Euler Equations. Flat-Terrain Case.}

\bigskip
\bigskip

\normalsize \rm

\centerline{\sc P. B{\'e}nard$^*$, R. Laprise$^\circ$, J. Vivoda$^+$, P. Smol\'{\i}kov\'a$^\dagger$ }
\bigskip
\bigskip
\centerline{\footnotesize $^*$ \sl Centre National de Recherches Météorologiques, Météo-France, Toulouse, France}
\medskip
\centerline{\footnotesize $^\circ$ \sl Université du Québec à Montréal, Montréal, Canada}
\medskip
\centerline{\footnotesize $^+$ \sl Slovak Hydro-Meteorological Institute, Bratislava, Slovakia}
\medskip
\centerline{\footnotesize $^\dagger$ \sl Czech Hydro-Meteorological Institute, Prague, Czech Republic}
\bigskip
\bigskip

\normalsize
\rm

\centerline{12 September 2003}

\bigskip
\bigskip
\bigskip
\bigskip

Corresponding address:

\medskip

Pierre Bénard

CNRM/GMAP

42, Avenue G. Coriolis

F-31057 TOULOUSE CEDEX

FRANCE

\bigskip

Telephone: +33 (0)5 61 07 84 63

Fax: +33 (0)5 61 07 84 53

e-mail: pierre.benard@meteo.fr

\newpage 
\centerline {ABSTRACT}
\bigskip

The stability of Semi-Implicit schemes for the Hydrostatic Primitive
Equations system has been studied extensively over the past
twenty years, since this temporal scheme and this system
represented a standard for NWP.
However, with the increase of computational power,
the relaxation of the 
hydrostatic approximation through the use of 
nonhydrostatic fully elastic systems is 
now emerging for future NWP as an attractive,
solution valid at any scale. 
In this context, several models employing 
the so-called Euler Equations together with 
a constant-coefficients semi-implicit time discretisation
have already been developed, but no solid justification
for the suitability of this algorithmic combination
has been presented so far, especially from the
point of view of robustness.
 
The aim of this paper is to investigate the 
response of this system/scheme in terms of stability
in presence of explicitly treated residual terms,
as it inevitably occurs in the reality of NWP.
This sudy is restricted to the impact of thermal
and baric residual terms (metric residual terms
linked to the orography are not considered here).
It is shown that conversely to what occurs 
with Hydrostatic Primitive Equations, 
the choice of the prognostic variables used 
to solve the system in time is of primary importance
for the robustness with Euler Equations. 
For an optimal choice of prognostic
variables, unconditionnally stable schemes can be obtained
(with respect to the length of the time-step),
but only for a smaller range of reference states than
in the case of Hydrostatic Primitive Equations.
This study also indicates that: (i) vertical coordinates
based on geometrical height and on mass 
behave similarly in terms
of stability for the problems examined here, and (ii)
hybrid coordinates induce an intrinsic instability, the
practical importance of which is however not completely
elucidated in the theoretical context of this paper.

\bigskip
\bigskip
\bigskip

\normalsize

\newpage

\section{Introduction}

In their most general definition, semi-implicit (SI) schemes 
consist in an arbitrary separation of the evolution terms of 
any dynamical system between some linear terms, treated 
implicitly, and non-linear residuals (NL residuals 
hereafter), treated explicitly. 
In meteorology, depending on the nature of the implicitly-treated
terms, three main types of SI schemes can be distinguished. The
coefficients of these linear terms may be: (i) constant both in time
and horizontally; (ii) constant in time only; and (iii) non-constant.

The first approach was initially introduced for meterological 
applications by Robert et al., 1972, and has been extensively
used in numerical weather prediction (NWP) since the solution 
of the resulting implicit system requires only basic 
techniques. However, due to large NL residuals, the stability
of these schemes is not formally guaranteed, especially for 
long time-steps.

The second and third approaches require more sophisticated 
techniques for solving the resulting implicit system, but 
they allow a significant reduction of the magnitude of 
the explicitly treated residuals, and hence, a potentially
better stability.

In the present paper, the terms "constant-coefficients 
SI schemes" exclusively refer to the above first category 
of SI schemes, and only these schemes are considered in 
all the following, unless expressly mentionned.


Historically, SI schemes have been first applied in NWP 
for solving the hydrostatic primitive equations (HPE) 
system, and extensive stability studies have 
been carried out with this system.
Simmons {\sl et al.}, 1978 (hereafter SHB78) 
investigated the practical stability of HPEs with the
three-time-level (herafter referred to as 3-TL) 
leap-frog constant-coefficients SI scheme in the terrain-following
pressure-based $\sigma$ coordinate.
To do so, they examined the effect of the leading 
NL residual terms on the stability when the SI 
reference temperature deviates from the actual 
temperature. 
They showed that, in the particular case where the 
complete model operator and the linearized SI 
operator have the same eigenfunctions,
the stability can be studied analytically.
In the more general case, when this latter 
condition is not fulfilled, Coté et al., 1983
(hereafter CBS83) showed that the stability 
can still be assessed, but at the price of a
"numerical analysis" which can be performed 
only in the space-discretized context;
the stability analysis then becomes 
an eigenvalue problem in a generalized 
state-vector space where the whole 
space- and time-discretized model
acts as a so-called "amplification matrix".
The salient result of SHB78 was that a
warm isothermal choice for the SI reference 
temperature profile resulted in a more stable
scheme than when using climatological profiles
for the SI reference-state, a rule which has been
widely followed in practical NWP applications. 
The two methods proposed by SHB78 and CBS83 
(analysis in simplified cases, and numerical 
analysis in the general case) have been adopted
by most of subsequent studies on the SI scheme 
stability.
CBS83 showed that, for the finite-element
vertical discretization of a 3-TL HPE model in 
$\sigma$ coordinate, the SI reference-state 
static stability had to be larger
than half the actual one, in order to achieve stability,
which explains and generalizes the previous 
results by SHB78.
Simmons and Temperton (1997, hereafter ST97) have extended the 
study of SHB78 to extrapolating two-time-level
(hereafter 2-TL) SI schemes, still in the HPE
system but for the more general hybrid-pressure
terrain-following $\eta$ coordinate
(in $\eta$ coordinate, the surface pressure in the 
SI reference-state also has to be considered for
the stability of the SI scheme).

However, with the increasing resolutions allowed by faster computers, 
nonhydrostatic NH models are now accessible to NWP, thus avoiding the
limitations associated with the hydrostatic assumption. However,
the aim for NWP models should be to achieve the relaxation of
the hydrostatic approximation while keeping the same
degree of efficiency as the former HPE SI semi-Lagrangian systems.
This pleads in favor of attempting to extend the SI 
technique to the new generation of nonhydrostatic semi-Lagrangian models.
Among the possible NH systems, the fully elastic
Euler Equations (EE) system is generally advocated
for several reasons, including
the possibility to build a "universal" atmospheric model
which dynamical kernel is valid for all scales used by the
meteorological community.

Several models using the EE system with a constant-coefficients SI scheme
have been recently developed (Tanguay {\sl et al.}, 1990;  
Laprise {\sl et al.}, 1995;  
Semazzi {\sl et al.}, 1995;  
Bubnov{\'a} {\sl et al.}, 1995, hereafter BHBG95; 
Caya and Laprise, 1999),
but the stability of such a combination
has not yet been studied in detail. 
Tanguay {\sl et al.} (1990) examined the stability of the
tangent-linear version of their model around the SI 
reference state (i.e. in the absence of non-linear terms).
Not surprisingly, they found that the system is 
unconditionally stable, since this follows from a general
property of any purely linear SI physical system.
However, the experience accumulated with former HPE systems clearly
indicates that some care must be taken when (explicitly treated) nonlinear
terms are present. Moreover, there is no objective reason
to expect an intrinsically more robust behavior of the
SI EE system compared to the SI HPE system. Especially, when
increasing the resolution to mesoscales, terms associated with
the orographic forcing or with physical processes could
have an increasingly stringent impact on the stability,
due to their increased contribution in the total evolution.
In BHBG95, we reported an unstable behavior for the EE system
with the SI time discretization, and this problem was
solved in the Eulerian context through the iteration 
of a part of the NL residual terms.
However the unstable behavior reappeared
when the resolution was refined or when the timestep 
was increased, as allowed by the implementation 
of a semi-Lagrangian scheme. 
This prompted us to undertake this study.
A general formalism for studying the stability of various 
time discretizations (including the SI scheme) and equation systems
(including the EE system) has been presented in Bénard (2003), 
B03 herefater.
This general method is applied here to the case of the 3-TL SI 
scheme for the EE system  with flat orography, in order to show
the essential role played by the choice of prognostic variables 
in the robustness of the system.

\section{Framework of analyses}
\label{sec_fram}

\noindent In this paper (except in Appendix B), the EE are cast in the pure 
unstretched terrain-following coordinate  $\sigma$ which can be 
 classically derived from the mass-based hydrostatic-pressure
coordinate $\pi$ (Laprise, 1992) through 
$\sigma = (\pi/\pi_s)$, where $\pi_s$ is the hydrostatic
surface pressure.
The $\sigma$ coordinate examined here is a particular case
of the general stretched hybrid-pressure 
terrain-following coordinate $\eta$ of BHBG95. 
However, the use of the $\sigma$ coordinate is advantageous for
the theoretical analysis since it possesses a much simpler
vertical metrics.
Starting from the general $\eta$ formalism, the equations for the 
$\sigma$ coordinate can be obtained by setting the arbitrary 
$A(\eta)$ and $B(\eta)$ functions in equation (9) of BHBG95 as follow:

\vspace{-0.4in} 
\begin{eqnarray}
A(\eta) & = &  0 \\
B(\eta) & = & \eta \equiv \sigma
\end{eqnarray}

\noindent In all the discussions of this paper, the flow 
will be assumed adiabatic inviscid and frictionless
in a non-rotating perfect-gas dry 
atmosphere with a Cartesian coordinate system. 
In these conditions, the complete set of Euler equations
can be easily derived from BHBG95 equations (1)-(8) with 
the same standard notations (for non-standard notations 
see Appendix A). The system in $\sigma$ coordinate writes:

\vspace{-0.4in} 

\begin{eqnarray}
\frac{d {\bf V}}{dt} + R T \frac{\bnabla p}{p}
+ \frac{1}{\pi_s} \frac{\dr p}{\dr \sigma} \bnabla \phi & = & 0
\label{eq_gene_1}\\
\frac{dw}{dt} +g \lp 1 - \frac{1}{\pi_s} \frac{\dr p}{\dr \sigma} \rp & = & 0 \\
\frac{dT}{dt} - \frac{R T}{C_{v}} . D_3 & = & 0 \\
\frac{dp}{dt} + \frac{C_{p}}{C_{v}} p D_3 & = & 0 \\
\frac{\dr q}{\dr t} + \int_0^1{ \lp {\bf V} \bnabla q + \bnabla {\bf V} \rp } d\sigma & = & 0 
\label{eq_gene_5}
\end{eqnarray}

\noindent where the three-dimensional divergence is given by:
\vspace{-0.4in} 

\begin{equation}
D_3 = \bnabla .{\bf V}
+  \frac{p}{\pi_s R T}  \lp \frac{\dr {\bf V}}{\dr \sigma} \rp \bnabla \phi 
-  g \frac{p}{\pi_s R T} \frac{\dr w}{\dr \sigma}
\end{equation}

\noindent and the geopotential horizontal gradient is:
\vspace{-0.4in} 

\begin{equation}
\bnabla \phi = \bnabla \phi_s
+  R \int_{\sigma}^1{\bnabla \lp \frac{\pi_s T}{p} \rp } d\sigma 
\end{equation}

\noindent We note $q=\ln(\pi_s)$, $p$ is the true pressure, and $w$ is the vertical 
velocity.

\noindent The theoretical analysis of the stability of non-linear systems
such as (\ref{eq_gene_1})-(\ref{eq_gene_5}) 
is not possible in the most general conditions and requires some 
simplifications in order to becomes algebraically tractable. 
For the analyses presented here, we use the same method 
and notations as in B03, which is basically a 
generalisation of the method proposed by SHB78.

\noindent In symbolic notations, the equations of the system to
be solved can be written as:

\vspace{-0.4in} 

\begin{equation}
\frac{\dr {\cal X}}{\dr t}= {\cal M}({\cal X})
\end{equation}

\noindent where ${\cal X}$ is the state variable, and ${\cal M}$ is the 
full model operator containing only spatial dependencies on ${\cal X}$.
For the definition of the SI scheme, as usually done in NWP, 
a SI reference-state $\cxst$ is chosen, and the system 
${\cal M}$ is linearized around $\cxst$,
resulting in a linear operator noted $\clst$.
The 3-TL SI time discretization writes:

\vspace{-0.4in} 

\begin{equation}
\frac{{\cal X}^+ - {\cal X}^-}{2 \Delta t} = \lc {\cal M}({\cal X}^0) - \clst.{\cal X}^0 \rc
            + \clst. \lp \frac{{\cal X}^+ + {\cal X}^-}{2} \rp                           
\end{equation}

\noindent where the superscript (-, 0, +) indicates variables at time
$(t- \Delta t)$, $t$, and  $(t+ \Delta t)$ respectively, where $\Delta t$ 
is the lenth of the timestep. 
The stability of the model is then conditioned by the structure of the
NL residual $({\cal M} - \clst)$. 
\medskip

For the purpose of analyses, the flow is assumed to consist of
small perturbations around a steady basic-state $\cxba$. Hence
${\cal M}(\cxba) = (\dr \cxba / \dr t) =0$, and the full model 
evolution ${\cal M}$ can be described by $\clba$, the linear-tangent 
operator of ${\cal M}$ around $\cxba$. 

The 3-TL SI time-discretized 
equation then becomes:

\vspace{-0.4in} 

\begin{equation}
\frac{{\cal X}^+ - {\cal X}^-}{2 \Delta t} = \lp \clba - \clst \rp .{\cal X}^0 
            + \clst. \lp \frac{{\cal X}^+ + {\cal X}^-}{2} \rp 
\label{eq_symb_SI}                            
\end{equation}

\noindent which represents the equation to be solved in the 
simplified framework examined here. Note that $\cxba \neq \cxst$
and hence $\clba \neq \clst$ thus giving rise to explicit contributions 
in (\ref{eq_symb_SI} ) 

\noindent The other simplifications adopted for the present analyses 
(as well as in SHB78), consist in assuming that both basic- 
and SI reference-states are  resting, isothermal and 
hydrostatically-balanced, with a uniform (plateau) orography.

As a consequence, in hydrostatic-pressure based coordinate, 
the $\cxba$ and  $\cxst$ states are simply characterized by their
uniform temperature and hydrostatic surface-pressure fields, i.e. by
the pairs of numbers ($\tba$, $\pisb$) and ($T^*$, $\pi_s^*$)
respectively. 
\noindent Finally, the domain is assumed two-dimensional 
(in a vertical plane) with $x$ as horizontal coordinate.

\medskip

\noindent The fact to examine the stability of the numerical
system in this highly simplified context of course leads to an
overestimation of the stability compared to what can be expected
for more complex flows with a fully non-linear model.
However if a scheme is found unstable here, it will have little
chances to be applicable in practice.
\medskip

\section{Impact of the SI-reference pressure} 
\label{sec_impP}

\noindent In this section, we examine the SI 
scheme for EE when the actual hydrostatic surface-pressure 
$\overline{\pi_s}$  
deviates from its SI reference counterpart $\pi_s^*$, 
everything else being equal for the two states. 
Hence, in this section the basic $\cxba$ and reference  $\cxst$ states
are characterized by ($\tba \equiv T^*$, $\pisb$) and  ($T^*$, $\pi_s^*$) respectively.
The linearized equations are derived  from (\ref{eq_gene_1})-(\ref{eq_gene_5}) or
from Eq. (13)-(21) of BHBG95, with the following set of prognostic variables:
horizontal divergence $D$, temperature $T$, $q = \ln(\pi_s)$,
and the following two nonhydrostatic variables as in BHBG95:

\vspace{-0.4in} 

\begin{eqnarray}
\hpca  & =  & \frac{p - \pi}{\pi^*} \\
\hd  & =  & -g \frac{\pi^*}{ m^* R T^*}\frac{\dr w}{\dr \eta}
\end{eqnarray}

\noindent where $\pi^*(\eta)$ is the
SI reference pressure profile, and $m^*(\eta)=(d \pi^* / d \eta)$. 
In $\sigma$ coordinate, these variables become:
\vspace{-0.4in} 

\begin{eqnarray}
\hpca  & =  & \frac{p - \pi}{\sigma \pi_s^*} \\
\hd    & =  & -g \frac{\sigma}{R T^*}\frac{\dr w}{\dr \sigma}
\label{eq_hd}
\end{eqnarray}

\noindent The $\clba$ system then writes:
\vspace{-0.4in} 

\begin{eqnarray}
\frac{\dr D}{\dr t} & = & - R {\cal G} \nabla^2 T 
   +  R T^*  \lc \frac{\pi_s^*}{\pisb} \rc({\cal G} - {\cal I} ) \nabla^2 \hpca 
   - RT^* \nabla^2 q \\
\frac{\dr \hd}{\dr t} & = & 
   - \frac{g^2}{R T^*} \lc \frac{\pi_s^*}{\pisb} \rc \drti (\drti+{\cal I})  \hpca \\
\frac{\dr T}{\dr t} & = & - \frac{R T^*}{C_v} ( D  +  \hd  ) \\
\frac{\dr \hpca}{\dr t} & = & \lc \frac{\pisb}{\pi_s^*} \rc {\cal S}D 
    - \frac{C_p}{C_v} \lc \frac{\pisb}{\pi_s^*} \rc ( D + \hd )  \\
\frac{\dr q}{\dr t} & = & -  {\cal N}D   
\end{eqnarray}

\noindent where the notations follow BHBG95 (see also Appendix A for 
the meaning of vertical operators ${\cal I}$, ${\cal G}$, ${\cal S}$, ${\cal N}$
and $\drti$ in $\sigma$ coordinate). 

\noindent The SI linear model $\clst$ can be derived in a similar way
except that $\pi_s^*$ is used instead of $\pisb$:
it is then formally identical to $\clba$, except that the factors 
in brackets in the above system become equal to 1. 
As a consequence, the explicitly-treated NL residual model ($\clba-\clst)$ 
in (\ref{eq_symb_SI}) is non-zero, leading to a potential source of instability
resulting from the departure of $\pisb$ from $\pi_s^*$.

\noindent An alternative formulation can be obtained if the variable 
$\hpca$ is replaced by a new variable $\pca$ defined by:

\vspace{-0.4in} 

\begin{equation}
\pca  =  \frac{p - \pi}{\pi}
\end{equation}

\noindent The form of the full nonlinear model ${\cal M}$
is modified in such a way that all occurences of $\hpca$ in 
the RHS must be replaced by $(\pi/\pi^*) \pca$, and the pressure
departure equation becomes:

\vspace{-0.4in} 

\begin{equation}
\frac{d \pca}{d t} =  \frac{\pi^*}{\pi} \lc \frac{d \hpca}{d t} \rc + 
      \frac{\pi \pca}{\pi^*} \frac{d}{dt} \lp \frac{\pi^*}{\pi} \rp,
\label{eq_pca}
\end{equation}

\noindent where the bracketed term is the RHS of the original
prognostic $\hpca$ equation. The last term of (\ref{eq_pca})
is identically zero for the linearized $\clba$ and $\clst$ systems.
Moreover the factor $(\pi^* /\pi)$ writes $(\pi_s^* /\pisb)$
in the present linearized context in $\sigma$ coordinate.
As a consequence, the new $\clba$ system writes:

\vspace{-0.4in} 

\begin{eqnarray}
\frac{\dr D}{\dr t} & = & - R {\cal G} \nabla^2 T 
      +  R T^*  ({\cal G} - {\cal I} ) \nabla^2 \pca 
      - RT^* \nabla^2 q \\
\frac{\dr \hd}{\dr t} & = & - \frac{g^2}{R T^*} \drti ( \drti + {\cal I}) \pca \\
\frac{\dr T}{\dr t} & = & - \frac{R T^*}{C_v} ( D  +  \hd  ) \\
\frac{\dr \pca}{\dr t} & = & {\cal S}D - \frac{C_p}{C_v} ( D + \hd )  \\
\frac{\dr q}{\dr t} & = & -  {\cal N}D   
\end{eqnarray}

\noindent Since this system has no dependency on $\pi_s^*$, it is
obvious that the SI reference system $\clst$ will have exactly the same form
as $\clba$ for this set of variables. 
In fact, it can be seen that using the variables 
($q$, $\pca$), and the $\sigma$ coordinate allows to completely
eliminate the SI reference surface-pressure $\pi_s^*$ from the model
formulation, in opposition to what occurs when using $\pi_s$ and/or
$\hpca$ as prognostic variables and/or the general hybrid $\eta$ coordinate.
For the new variable $\pca$, there is no 
NL residual terms, and hence no potential source 
of instability due to the discrepancy between
$\pi_s$ and $\pi_s^*$ for the examined problem. 
As a direct consequence, no stability analysis 
is necessary here to conclude that the variable $\pca$
is better suited to the design of a SI scheme than $\hpca$.

\noindent Similar algebraic derivations show that for the
particular problem examined here, the various possible 
choices for the prognostic pressure variables fall into two 
classes:

\begin{list}{}{}
\item [](i) variables leading to potentially unstable SI: $p$, $p/p_{0}$, $p - \pi$, $p/ \pi^*$, $\hpca$
\item [](ii) variables leading to stable SI: $\ln(p)$, $\ln(p/p_{0})$, $p/ \pi$, $\ln(p / \pi)$, $\pca$
\end{list}

\noindent where $p_{0}$ is an arbitrary constant.
It is worth emphasizing that the above 
statement holds for height-based coordinates as well
as for the mass-based coordinate that was used here
(with of course the restriction due to the fact that
the variables involving $\pi$ are not natural with 
height-based coordinates). 
These properties follow immediately from
the derivation of the corresponding linear system
in the same context, for height-based coordinates.
From now on, the new variable
$\pca$ will be used instead of the 
original variable $\hpca$ used in BHBG95.

\section{Impact of the SI reference temperature}
\label{sec_impT}

\noindent In this section, we examine the stability of the SI 
scheme for EE with the 
prognostic variables ($D$, $\hd$, $T$, $\pca$, $q$), 
when the basic uniform temperature $\tba$ 
deviates from the SI reference state temperature
$T^*$, everything else being equal for the 
two states.
The variable $\hd$ is still given by (\ref{eq_hd}). 
As a consequence, the 3-dimensional divergence  $D_3$
(see Eq. (20) of BHBG95) writes for the $\clba$ system: 

\vspace{-0.4in} 

\begin{equation}
D_3 = D + \frac{T^*}{\tba} \hd,
\label{eq_D3hat}
\end{equation}

\noindent and the direct linearization of the original
system yields:
\vspace{-0.4in} 

\begin{eqnarray}
\frac{\dr D}{\dr t} & = & - R {\cal G} \nabla^2 T 
   +  R \tba ({\cal G} - {\cal I} ) \nabla^2 \pca 
   - R \tba \nabla^2 q 
   \label{eq_sys1D}\\
\frac{\dr \hd}{\dr t} & = & 
   - \frac{g^2}{R T^*}  \drti ( \drti + {\cal I}) \pca \\
\frac{\dr T}{\dr t} & = & - \frac{R \tba}{C_v} ( D  +  \frac{T^*}{\tba} \hd  ) \\
\frac{\dr \pca}{\dr t} & = &  {\cal S}D - \frac{C_p}{C_v}  ( D + \frac{T^*}{\tba} \hd )   
  \label{eq_sys1P}\\
\frac{\dr q}{\dr t} & = & -  {\cal N}D   
\end{eqnarray}

\noindent The $\clst$ operator is defined in a similar way, 
simply replacing $\tba$ by $T^*$ in the RHS of the above system.

\noindent The method for the stability analysis exactly follows 
the one proposed in B03, and 
the reader is invited to refer to this paper for more details 
on the notations and the algebraic developments.
The above system is first shown to fulfil the four conditions [C1]--[C4] 
required for making possible the space-continuous analyses
with the proposed method. 
The number of prognostic variable is $P=4$
in the sense of B03, and the space-continuous state-vector is
 ${\cal X}=({\cal X}_1, \ldots, {\cal X}_4)=(D, \hd, T, \pca)$.
The linear operator in [C1] involves $l_1= \drti$
applied to (\ref{eq_sys1D}) and $l_4= ( \drti + {\cal I})$
applied to (\ref{eq_sys1P}) as in section 7.1 of B03. 
The condition [C'2] requires $\tba > 0$, and
the normal modes of the system are then:

\vspace{-0.4in} 

\begin{equation}
{\cal X}_j (x,\sigma)  =  \cxha_j \; \exp(ikx)\, \sigma^{(i \nu - 1/2)}  \; , \;\;\; {\rm for}\; j \in (1, \ldots, 4)
\end{equation} 

\noindent where $(k,\nu) \in \Real$ (note that $\nu$ is a 
{\sl non-dimensional} vertical wavenumber).
In this particular case, the four components $(f_1, \ldots, f_4)$ 
of the  shape function $f$ introduced in B03 are identical. The vector
function $f$ represents the geometry of any normal mode of the
time and space-continuous system. 
The verification of [C3], [C4] proceeds easily, as in B03; 
for [C3], we have:

\vspace{-0.4in} 

\begin{eqnarray}
\xi_1 & = &  i \nu - 1/2\\
\xi_4 & = &  i \nu + 1/2,
\end{eqnarray}

\noindent and for [C4]:

\vspace{-0.4in} 

\begin{eqnarray}
\muba_{13} & = &  \must_{13} = - k^2 R \\
\muba_{14} & = &   k^2 R \tba (i \nu + 1/2) \; , \;\;\; \must_{14}  =    k^2 R T^* (i \nu + 1/2) \\
\muba_{24} & = &  \must_{24} =  (\nu^2 + 1/4)  \frac{g^2}{R T^*} \\
\muba_{31} & = &  - \frac{R \tba}{C_v}  \;      , \;\;\; \must_{31}  =  \muba_{32}  =  \must_{32} = - \frac{R T^*}{C_v}  \\
\muba_{41} & = & \must_{41} =  1 - \frac{C_p}{C_v} (i \nu + 1/2) \\
\muba_{42} & = &  - \frac{C_p}{C_v} (i \nu + 1/2) \frac{T^*}{\tba} \;  , \;\;\;  \must_{42}  =   - \frac{C_p}{C_v} (i \nu + 1/2).
\end{eqnarray}

\noindent For the stability analysis, the growth of any mode
with the shape function $f$ is examined. The analysis hence 
consists in  solving (\ref{eq_symb_SI}) assuming 
${\cal X}_{(t=-\Delta t)} = \cxha^- .f(x,\sigma)$ together with:
  
\vspace{-0.4in} 

\begin{eqnarray}
{\cal X}_{(t = 0)} & = & \lambda {\cal X}_{(t=-\Delta t)} \\
{\cal X}_{(t = \Delta t)} & = & \lambda^2 {\cal X}_{(t=-\Delta t)},
\end{eqnarray}

\noindent where the unknowns are the complex polarisation vector $\cxha^-$
and the numerical complex growth rate $\lambda$. 
As stated in B03, the 3-TL SI scheme is a particular ICI scheme
with $N_{\rm iter}=1$ and $\mu(\lambda) =2 - 1/\lambda$. 
Hence, in the formalism of B03, the stability problem reduces to:

\vspace{-0.4in} 

\begin{equation}
\lp
\begin{array}{ccc}
(2 \lambda - 1) I_4  & -I_4  &   0_4    \\
         M_1         &  M_2  &   M_3    \\
- \lambda I_4        &  0_4  &   I_4 
\end{array}
\rp . \cz  = {\bf M}. \cz = 0.
\label{eq_defm}
\end{equation} 

\noindent where the generalised state-vector $\cz$ is defined 
by $\cz=(\cxha^-,\cxha^{+(0)},\cxha^{+(1)})$, 
$I_4$ and $0_4$ are the unit and null $4$th-order 
matrices respectively. In the 3-TL SI framework, the sub-matrices 
$M_1$, $M_2$ $M_3$ are defined by:

\vspace{-0.4in} 

\begin{eqnarray}
\lp M_1 \rp_{ij} & = & - \delta_{ij} -   \Delta t \frac{\overline{\mu}_{ij}}{\xi_i} \\
\lp M_2 \rp_{ij} & = & - \Delta t \frac{1}{\xi_i} \lp \overline{\mu}_{ij} - \mu^*_{ij} \rp \\
\lp M_3 \rp_{ij} & = & + \delta_{ij}  -   \Delta t \frac{\mu^*_{ij}}{\xi_i} 
\end{eqnarray}

\noindent where all notations follow B03.
The possible values of $\lambda$ for the normal mode structure
that we examine, are thus given by the roots of
the following polynomial equation in $\lambda$:

\vspace{-0.4in} 

\begin{equation}
{\rm Det} ( {\bf M}) = 0
\label{eq_detm}
\end{equation}

\noindent In this simple case, the determinant is easily expanded 
algebraically, and yields:

\vspace{-0.4in} 

\begin{eqnarray}
\frac{(\xxm)^4}{\Delta t^4}
& + & c^2 \frac{(\xxm)^2}{\Delta t^2} (\xxp) \lp  k^2 \xxa +  n \nba \, \xxb \rp  
\nonumber \\
& + & k^2 N^2 c^2 (\xxp)^2 \lc \lp \frac{\xxa \xxb C_p - (\xxp)^2 R}{C_v} \rp
- nH \lp \xxa \xxb - (\xxp)^2 \rp  \rc = 0
\label{eq_stru2}
\end{eqnarray}

\noindent where the time-discretised response factors are defined by:

\vspace{-0.4in} 

\begin{eqnarray}
\xxm  & = &  \frac{\lambda^2-1}{2}  
\label{eq_def_xxp}\\
\xxp  & = &  \frac{\lambda^2+1}{2}  \\
\xxa  & = &  (\xxp) + \alpha \lambda  \\
\xxb  & = &  (\xxp) - \frac{\alpha}{1 + \alpha} \lambda,
\label{eq_def_xxb}
\end{eqnarray}

\noindent and:

\vspace{-0.4in} 

\begin{eqnarray}
\alpha & = &  \frac{\tba - T^*}{T^*}  \\
  c^2 & = & R T^* (C_p / C_v) \\
    H & = & R T^*/g \\
    N & = & g/ \sqrt{C_p T^*} 
    \label{eq_def_N}\\
    n & = & (i \nu + 1/2) H^{-1} \\
 \nba & = & (- i \nu + 1/2) H^{-1}
\end{eqnarray}

\noindent This eighth-degree complex polynomial equation in $\lambda$
can be solved numerically: for any  pair
($k$, $\nu$), the modulus of the eight roots $\lambda$ 
give the growth rate of the eight corresponding eigenmodes
(four physical modes and four computational modes, due
to the 3-TL discretization). If one of the roots has a
modulus larger than one, then the corresponding mode is
unstable. The stability of the scheme for the structure
function $f$ corresponding to a pair ($k$, $\nu$) is then 
given by the maximum modulus of the eigth corresponding 
eigenvalues:

\begin{equation}
\Gamma = {\rm Max} \lp \left | \lambda_i \right |\rp \; , \;\; i \in (1,\ldots,8)
\end{equation}

\noindent The criterion for unconditional stability 
of the scheme with respect to the time-step
can be found by requiring stability at the large
time-steps limit in the above equation. The terms containing
$\xxm$ are then vanishing and the equation for the 
growth rate becomes:

\vspace{-0.4in} 

\begin{equation}
(\xxp)^2 \lc \xxa \xxb \frac{C_p}{C_v} - (\xxp)^2 \frac{R}{C_v} 
- nH \lp \xxa \xxb - (\xxp)^2 \rp  \rc = 0
\label{eq_unc}
\end{equation}

\noindent The four modes represented by $(\xxp)^2 =0$ are always 
neutral. Conversely, in the other set of roots the short vertical
modes are always unstable. In effect, for these roots, substituting
the time discretization response factors by their value leads to:

\vspace{-0.4in} 

\begin{equation}
(1+ \alpha)(\lambda^2 + 1)^2 - 2 \alpha^2 \lambda (\lambda -1)^2
\lc i \nu  + (1/2 - C_p/C_v)\rc = 0
\label{eq_AGR1}
\end{equation}

\noindent For short modes, $\nu >> 1$,  hence
the modulus of the four roots becomes close to:

\vspace{-0.4in} 

\begin{equation}
\left | \lambda \right |_{(1,2,3,4)} \approx \frac{2 \alpha^2}{1 + \alpha} \nu
\end{equation}

\noindent and obviously this leads to large instabilities.
As a consequence, except in the degenerated 
case $\alpha=0$, the scheme cannot be unconditionally 
stable in $\Delta t$, because for large time-steps, short enough
vertical modes are unstable. 


\medskip

\noindent However, drawing on the results of the previous
section, the sensitivity of the stability to the choice 
of the prognostic variables  is suspected, and the 
relevance of the original choice $\hd$ could be questioned. 
A close inspection of the algebra in the above 
analysis indicates that the source
of the problem lies in the small discrepancy between 
$\xxa \xxb$ and $(\xxp)^2$ in (\ref{eq_unc}), 
which in turn is linked to the discrepancy between
the $D$ and $\hd$ factors in the RHS of (\ref{eq_D3hat}).
This suggests the use of an alternative variable $\nd$ 
which would be defined in the general hybrid
coordinate $\eta$ by:

\vspace{-0.4in} 

\begin{equation}
\nd   =   -g \frac{p}{ m R T} \frac{\dr w}{\dr \eta}
\label{eq_def_d}
\end{equation}

\noindent where $m=(\dr \pi /\dr \eta)$.
In the present linear and hydrostatically-balanced
context with $\sigma$ coordinate, $\nd$ simplifies to:

\vspace{-0.4in} 

\begin{equation}
\nd   =   -g \frac{\sigma}{ R \tba} \frac{\dr w}{\dr \sigma}
\end{equation}

\noindent and the linear 3-dimensional divergence  writes:

\vspace{-0.4in} 

\begin{equation}
D_3 = D + \nd
\end{equation}

\noindent The linear system $\clba$ becomes:
\vspace{-0.4in} 

\begin{eqnarray}
\frac{\dr D}{\dr t} & = & - R {\cal G} \nabla^2 T 
   +  R \tba ({\cal G} - {\cal I} ) \nabla^2 \pca 
   - R \tba \nabla^2 q 
   \label{eq_sys11D}\\
\frac{\dr \nd}{\dr t} & = & 
   - \frac{g^2}{R \tba}  \drti ( \drti + {\cal I}) \pca \\
\frac{\dr T}{\dr t} & = & - \frac{R \tba}{C_v} ( D  +  \nd  ) \\
\frac{\dr \pca}{\dr t} & = &  {\cal S}D - \frac{C_p}{C_v}  ( D + \nd )   
  \label{eq_sys11P}\\
\frac{\dr q}{\dr t} & = & -  {\cal N}D   
\end{eqnarray}

\noindent Since $\hd$ and $\nd$ have the same value in the
SI reference state, the general design of the SI scheme 
is unchanged: The linear system $\clst$ is formally identical
to the previous one, hence the modification does not
change the SI equation to be solved.
The stability analysis for this new system
can be done exactly in the same way as presented
above, and
the stability equation (\ref{eq_stru2}) for a given geometry 
($k$, $\nu$) becomes:

\vspace{-0.4in} 

\begin{equation}
\frac{(\xxm)^4}{\Delta t^4} 
+ c^2 \frac{(\xxm)^2}{\Delta t^2} (\xxp) \lp k^2 \xxa  
+  \,n \nba \xxb \rp  
+ k^2 N^2 c^2 (\xxp)^2 \, \xxa \xxb \,  = 0
\label{eq_stru3}
\end{equation}

\noindent As previously, the unconditional stability in $\Delta t$ can be examined by 
neglecting all terms containing $\xxm$, and the numerical growth rate is then
given by:

\vspace{-0.4in} 

\begin{equation}
(\lambda^2 +1)^2 
(\lambda^2 + 1 + 2 \alpha \lambda) 
(\lambda^2 + 1 - 2 \frac{\alpha}{1+ \alpha} \lambda)  = 0
\label{eq_AGR2}
\end{equation}

\noindent The four modes involved by the first factor are
always neutral. Basic algebraic manipulations show that
the modes involved in the second factor are stable for
$-1 \leq \alpha \leq 1$, while the last factor requires
$-1/2 \leq \alpha$ for stability. Finally, considering 
the definition of $\alpha$, the scheme is unconditionally 
stable in $\Delta t$ when the following condition 
is fulfilled:

\vspace{-0.4in} 

\begin{equation}
\frac{T^*}{2} \leq \tba \leq 2 T^*
\label{eq_criter_EE}
\end{equation}

The above type of stability analyses can be performed in the same way 
for any other vertical-velocity related prognostic variable. 
From the set $\la w \right.$, $(\dr w / \dr \eta)$, $\hd$, $ \left. \nd \ra$,
only the last variable is then found to allow a non-vanishing range of 
unconditional stability in mass-based coordinates.
But conversely to what occurred in the previous section,
this result is now dependent on the type of vertical 
coordinate used: for height-based coordinates, the same kind 
of analysis (see Appendix B) shows that the two most natural choices 
$\la w \right.$, $\left. \dr w / \dr z  \ra$ lead to the same unconditional
stability criterion than one obtained here for $\nd$ variable
in mass-based coordinate.
Hence, from the point of view of the asymptotic stability
at long time-steps, height-based and mass-based coordinates 
behave identically provided "optimal" variables are chosen.
The fact that $\nd$ in mass-based coordinates 
and $(\dr w / \dr z)$ in 
height-based coordinates behave similarly can be 
understood intuitively since these two variables are
in fact two expressions of a same concept in the 
present simplified context (i.e. they both represent 
the true vertical divergence here).
The reason why $w$ is stable in height-based coordinates
but unstable in mass-based coordinates is more subtle:
as suggested by the above analyses, the robustness of the
SI scheme appears to be highly correlated to the existence
of NL residuals in the elastic term $D_3$; when $w$ is used
as a prognostic variable in height-based coordinates, $D_3$ 
can be readily obtained from $D + ( \dr w / \dr z)$, and 
has no NL residual, thus leading to a stable 
scheme in the examined context.
With mass-based coordinates however, if $w$ is chosen as 
a prognostic variable, $D_3$ must be evaluated through
$D - (g p/ RT)( \dr w / \dr \pi)$ and the existence of
a NL residual when $T \neq T^*$ may lead to instabilities.
The fact that the vertical divergence $\nd$
is completely imposed as a prognostic variable if a robust
SI scheme is desired for the EE system with flat-terrain
is thus a specificity of  mass-based coordinates.

Another important result from the above analysis is that,
even with an optimal choice of the prognostic variables,
the stability range is dramatically 
reduced for EE system compared to the HPE system, whatever
vertical coordinate is used. For HPEs (see e.g. SHB78),
the criterion (\ref{eq_criter_EE}) would write:

\vspace{-0.4in} 

\begin{equation}
0 \leq \tba \leq 2 T^*
\label{eq_criter_HPE}
\end{equation}

\noindent Chosing a very warm $T^*$ garantees stability in HPEs 
while this is no longer the case for EEs. Moreover, if the 
actual temperature were to vary by more than a factor 
4 in the atmosphere, the SI technique examined 
here could not offer any stable scheme for the EE  system.
The SI EE system is thus less stable {\sl by nature}
than the SI HPE system, and its applicability for NWP
is only made possible thanks to the moderate 
variability of the thermal field in the
terrestrial atmosphere.

\section{Practical Implications}
\label{sec_pract}

\begin{figure}[p]
\epsfxsize=\figwidth
\centerline{\epsfbox{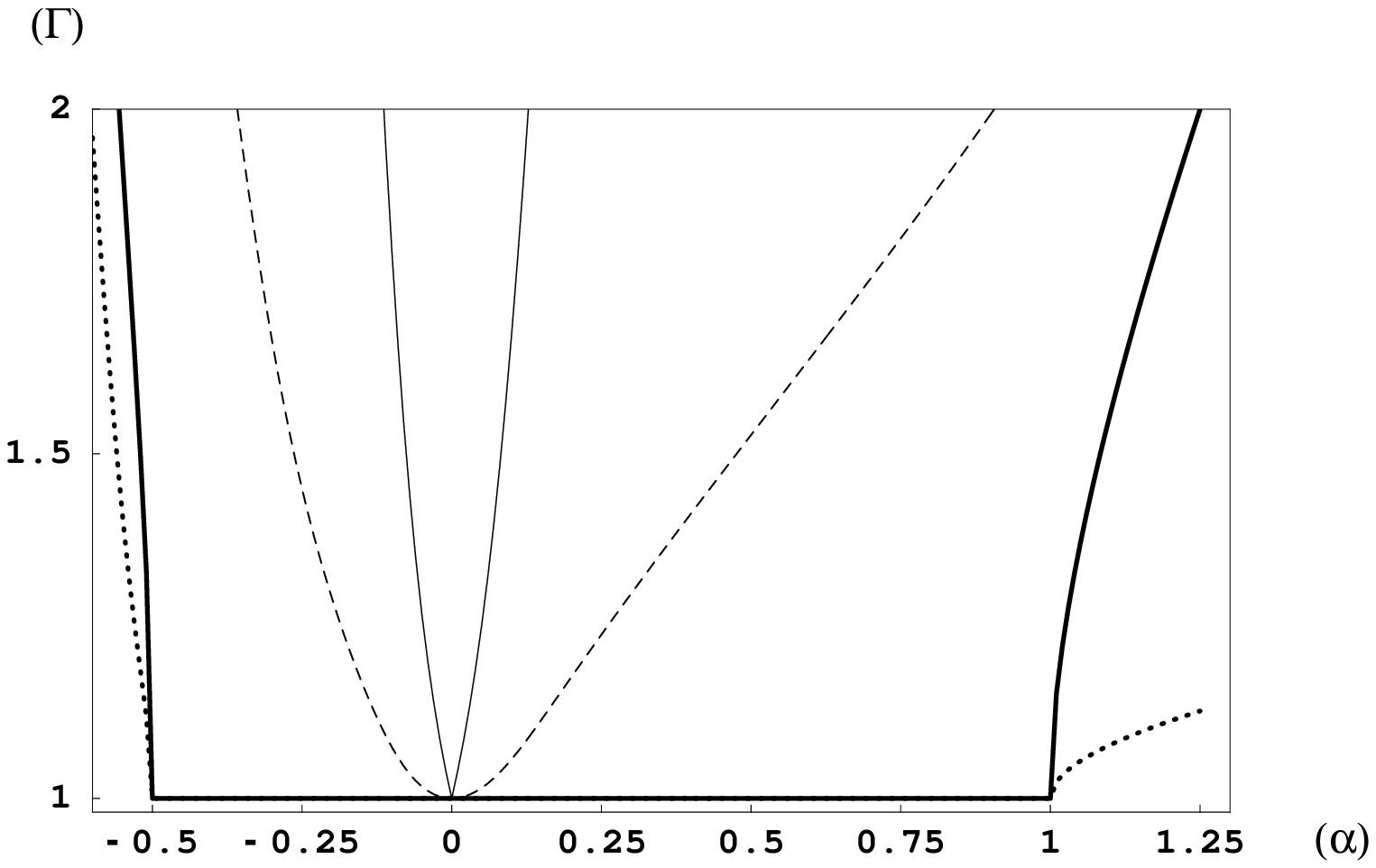}}
\caption{\label{fig_GR_anal} Analytic growth rates (GR) 
 for the simplified problem as a function of the 
 nonlinearity parameter $\alpha$.
  Thick line: asymptotic GR for variable $\nd$, 
  solid line: asymptotic GR for variable $\hd$,
  dotted line: practical GR for variable $\nd$.
  dashed line: practical GR for variable $\hd$
  (see text for the meaning of practical GR).
  }
\end{figure}

For a given mode, the root of maximum modulus
in (\ref{eq_AGR1}) and (\ref{eq_AGR2}) gives the asymptotic
growth rate for large timesteps for the two variables $\hd$ and $\nd$
examined above. For the variable $\hd$, this 
asymptotic growth rate is function of $\alpha$ and 
$\nu$, hence the instability is  directly linked to the 
vertical resolution of the model; for the variable
$\nd$ the asymptotic growth rate is function of
$\alpha$ only.
Figure \ref{fig_GR_anal} shows the asymptotic growth rate 
for a vertical spacing of 100 m (i.e. a minimum vertical 
wavelength of 200 m), for $\hd$ and $\nd$, in the conditions of
the above analyses. 
Assuming a typical value $\pm 0.25$ for $\alpha$
in realistic conditions, the range of growth rate for 
the variable $\hd$ is clearly incompatible with a stable 
integration of the SI scheme with very long time-steps.
In practical applications however, 
the time-step is bounded, and the growth,
if present, may not endanger
significantly the stability of the scheme. 
However, direct numerical solution of (\ref{eq_stru2})
shows that in fact any mode ($k$, $\nu$) 
is unstable for any timestep, when 
$\alpha \neq 0$. 
For instance, in Fig. \ref{fig_GR_anal} are also plotted
the numerical growth rates obtained 
for a mode which geometry is close to
the shortest mode for a typical mesoscale future NWP 
limited-area target configuration with 
3-TL semi-Lagrangian scheme 
($\Delta x$ = 2000 m, $\Delta z$ = 100 m, and
$\Delta t$ = 20 s): the system with $\hd$  variable
is significantly unstable as soon as $\alpha \neq 0$, 
even for this modest timestep. 
The $\hd$ variable is clearly demonstrated as not suitable 
for use in a SI EE scheme in mass-based coordinates.
This was one of the reasons why an iterative procedure
was required and applied in the NH model described 
in BHBG95.

An additional remark concerning the stability for finite
timesteps is that surprisingly, when an optimal variable 
is chosen in height-based coordinates, the SI scheme
which is stable for long timesteps 
becomes slightly unstable for finite timesteps
as soon as $\alpha \neq 0$, as seen in Appendix B. 
A similar behaviour was found in B03 for the 
one dimensional acoustic system with long time-steps
in height-based coordinates.
This unstable behaviour can be interpreted as originating 
from the fact that in height-based coordinates, the time-continuous 
normal modes of the $\clba$ system are not normal modes of 
$\clst$, as pointed out in Appendix B.
As a consequence, the time-continuous evolution of any
normal mode of $\clba$ by the $\clst$ system contains
a growing component (i.e. a complex non-real frequency) 
as soon as $\alpha \neq 0$. 
The SI scheme is then not able to ensure a stable 
evolution for some of these components.
This type of instability does not occur for mass-based
coordinates, as seen in Fig. \ref{fig_GR_anal}.
From the theoretical point of view, this behaviour
is in defavour of height-based coordinates, but it 
should be noted that the resulting instabilities 
are generally moderate (not shown) and could 
be controlled by the use of slightly
damping algorithms (e.g. SI decentering as 
in Laprise, 1995) or by the diffusive processes
acting in a complete model.

\medskip
The stability of the SI scheme can be studied
for non-isothermal linear atmospheric flows,
using the "numerical analysis" method
proposed by CBS83: The linearized equations have first
to be vertically discretized. For a given 
eigenmode of the horizontal $\nabla^2$ operator,
the actual and reference model operators can 
then be expressed as two matrices 
$\overline{\bf L}$ and ${\bf L}^*$ operating on
the perturbation state-vector {\bf X}.
The equation of model evolution (\ref{eq_symb_SI}) 
then writes:

\vspace{-0.4in} 

\begin{equation}
{\bf Y}^{t+ \Delta t} = {\bf A}.{\bf Y}^{t}
\label{eq_matr}
\end{equation}

\noindent where the generalized state vector ${\bf Y}^{t}$
is defined by $({\bf X}^{t},{\bf X}^{t- \Delta t})$,
and the amplification matrix {\bf A} is given by:

\vspace{-0.4in} 

\begin{equation}
 {\bf A} = \left ( 
 \begin{array}{cc}
   2 \Delta t \lp {\bf I_X} - \Delta t {\bf L}^* \rp^{-1} \!\!\!\!\!
     .  \lp \overline{\bf L} - {\bf L}^*\rp
 &  \lp {\bf I_X} - \Delta t {\bf L}^* \rp^{-1} \!\!\!\!\!
    .   \lp {\bf I_X} + \Delta t {\bf L}^* \rp \\
   {\bf I_X} & {\bf 0_X}
 \end{array} \right )
\end{equation}

\noindent where ${\bf I_X}$ and  ${\bf 0_X}$ are the identity 
and null operators in the state-vector space respectively.
Noting  $\Gamma$ the largest modulus of {\bf A}
eigenvalues, the scheme is stable if 
$\Gamma \leq 1$ and unstable if $\Gamma > 1$.
The eigenvector associated with $\Gamma$
gives the vertical structure and polarisation of the most 
unstable mode.

\begin{figure}[p]
\epsfxsize=\figwidth
\centerline{\epsfbox{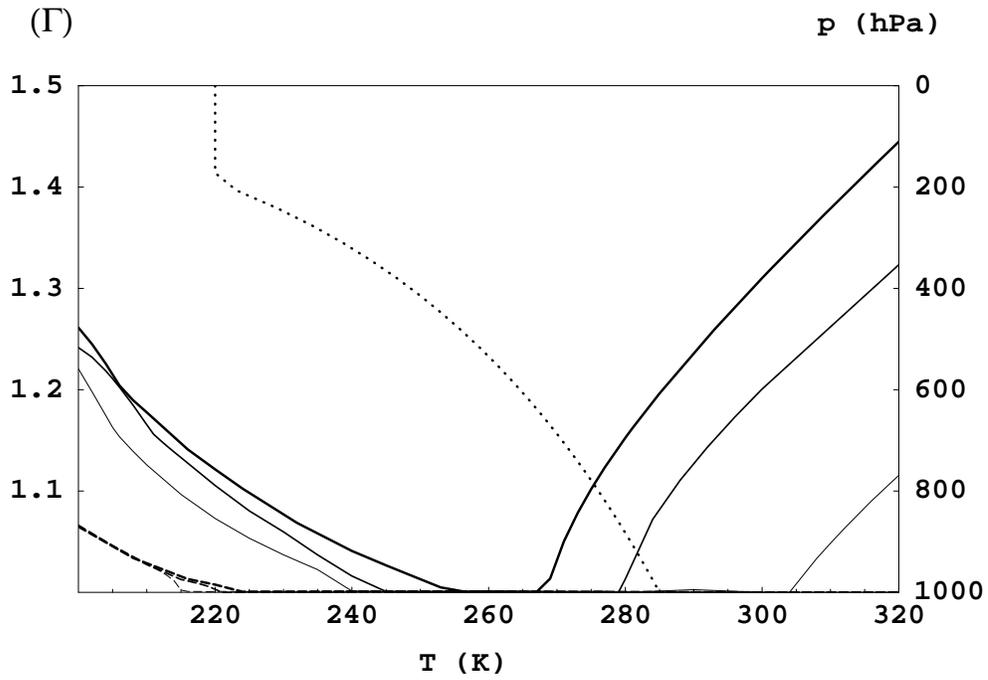}}
\caption{\label{fig_GR_prof} Numerical growth rate  
 (left axis) for the realistic thermal profile 
 (dotted line - pressure on right axis),as a function of $T^*$. 
 From thin to thick lines: 10, 20, 30 levels.
  solid line: variable $\hd$,
  dashed line: variable $\nd$.
  }
\end{figure}

\medskip
Our implementation of this "numerical analysis" method has been validated
by comparison with the results of the above
analyses: the agreement was found very good
(not shown).
Here, the "numerical analysis" method is applied to quantify the impact of 
the change from $\hd$ to $\nd$
in more realistic situations than the one 
assumed in the above analyses.
Following SHB78, a realistic actual temperature 
profile is chosen: it consists in a tropospheric 
profile with a quasi-uniform dry static stability and 
an isothermal stratosphere above 200 hPa, 
as depicted in Fig. \ref{fig_GR_prof} 
(the surface temperature is set to 285 K). 
The vertical discretization of $\clba$ and 
$\clst$ models follows BHBG95. 
The numerical analysis is performed for 
the same typical future NWP target values for
($\Delta x$, $\Delta t$) as in 
Fig. \ref{fig_GR_anal}.
The stability is examined for values of $T^*$
varying in the interval [200 K, 320 K].

Figure \ref{fig_GR_prof} shows the growth
rates obtained for 10, 20 and 30 
regularly-spaced $\sigma$ levels, for the
$\hd$ and $\nd$ variables. The theoretical
disadvantage of $\hd$ is confirmed:
for high vertical resolutions, the scheme 
is unstable for almost every value of $T*$, with
growth rates uncompatible with a practical
use.  The logarithm of the growth rate is found roughly
proportional to the horizontal wavenumber
$k$, confirming that troubles linked to the 
presence of NL terms become more and more stringent 
when the horizontal resolution is increased.
Moreover, reducing the timestep is found 
to be of no help for a given forecast range
because the logarithm of the growth rate is 
always roughly proportional to the time-step.
Conversely, the use of variable $\nd$ 
results in a stable scheme provided 
$T^*$ is chosen warm enough in the examined
interval (and up to 440 K here, consistenly 
with analyses).

\begin{figure}[p]
\epsfxsize=\figwidth
\centerline{\epsfbox{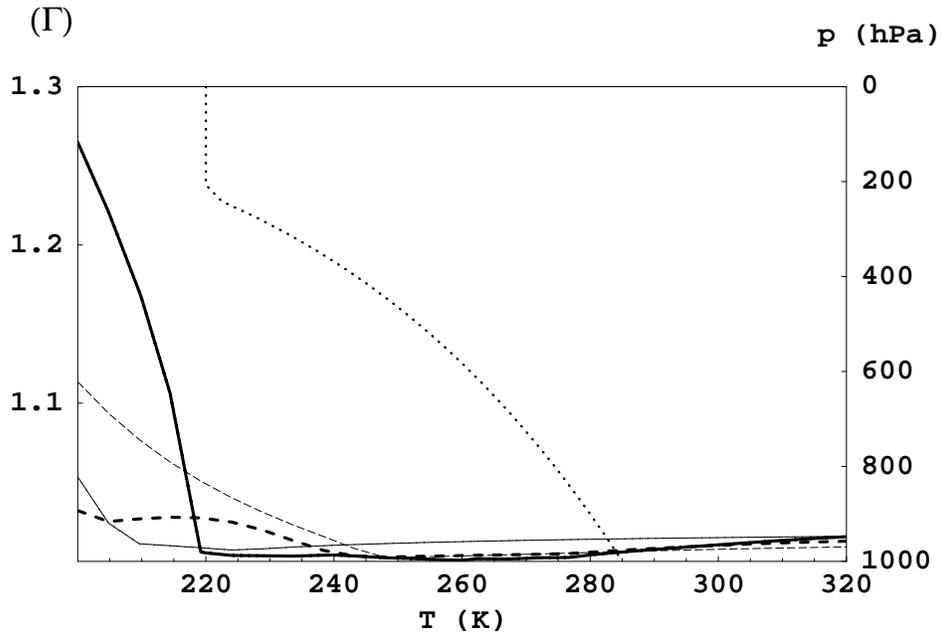}}
\caption{\label{fig_GR_pres} Numerical growth rate  
 (left axis) for the realistic thermal profile 
 (dotted line - pressure on right axis),as a function of $T^*$. 
  Thick lines: $\pi_s^*$ = 1213.25 hPa,
  thin lines : $\pi_s^*$ = 813.25 hPa,
  solid line: variable $\hpca$ with $\sigma$ coordinate,
  dashed line: variable $\pca$ with $\eta$ coordinate.
  }
\end{figure}

Taking $\nd$ as prognostic variable, the
impact of the choice of the nonhydrostatic pressure
variable ($\hpca$ vs. $\pca$) can be examined 
numerically, still in the same context. It has been
shown in section \ref{sec_impP} why the system could be potentially unstable when
$\pisb$ deviates from $\pi_s^*$. 
To confirm and illustrate this statement, two values of $\pi_s^*$
are chosen (813.25 hPa, and 1213.25 hPa), while
$\pisb$ is assumed to be 1013.25 hPa, still for the 
same thermal profile and experimental 
context as above, and for 30 regularly-spaced
$\sigma$-levels. 
The solid lines in Figure \ref{fig_GR_pres} show
the growth rates obtained for the variable $\hpca$
(as explained in section \ref{sec_impP}, the 
growth rate has no dependency 
on $\pi_s^*$ when the variable $\pca$ is used,
hence the corresponding growth rate 
for the variable $\pca$ can be seen in 
Fig. \ref{fig_GR_prof}, regardless of the 
value of $\pi_s^*$). 
The use of the $\hpca$ variable induces a global
decrease in stability especially for $\pi_s^*$=1213.25 hPa
and low values of $T^*$.
For high values of $T^*$, an instability 
subsists for both values of $\pi_s^*$, but
it is weak (growth rates around 1.015), and
the practical risk of using $\hpca$ for 
NWP is hence difficult to assess from this 
simplified theory and should be evaluated 
in a more realistic context. However, since 
there is no other potential advantage of
using $\hpca$ instead of $\pca$, this 
question is probably not of primary interest.

Finally, examination of the above algebra shows that
the hybrid $\eta$ coordinate can also be 
responsible for instabilities when 
$\pisb$ deviates from $\pi_s^*$, as it may e.g. be the
case when the terrain is uniformly elevated.
This is linked
to the fact that in $\eta$ coordinate, 
all vertical operators (${\cal G}$, ${\cal S}$,
${\cal N}$, $\drti$) deviate from
their SI-reference counterpart when 
$\pisb \neq \pi_s^*$, the resulting NL 
residuals potentially leading to instability. 
The discrepancy between the vertical metrics
of $\clba$ and  $\clst$ models makes impossible 
the analysis for a general choice of $A$ and 
$B$ functions; hence, we use the "numerical analysis"
method proposed by CBS83 to examine the impact
of using a hybrid coordinate instead of a pure
terrain-following coordinate when $\pisb \neq \pi_s^*$.

The dashed lines in Figure \ref{fig_GR_pres}
show the growth rates obtained for the variable
$\pca$ when a regularly flattening unstretched
$\eta$ coordinate, defined by:
$\pi(\eta)= - \rho \eta \ln(\eta) p_{00} + 
\eta \lc 1 + \rho \ln(\eta) \rc \pi_s$
(with $p_{00}$=1013.25 hPa, and $\rho$=0.25) is used
with the same realistic thermal profile and
discrepancies of $\pm 200$ hPa between
$\pisb$ and $\pi_s^*$ as above.
For high values of $T^*$, the loss of 
stability is of similar magnitude 
as the one resulting from the choice of $\hpca$
as prognostic variable. 
It is noteworthy that the instability linked
to the use of the hybrid coordinate in EE cannot
be eliminated by choosing high values for
$T^*$, as it was the case for HPE system 
(Simmons and Temperton, 1997), hence there is indication that
hybrid $\eta$ coordinate possesses a slight 
theoretical disadvantage from this point of 
view, but this may be not redhibitory in practice.
In other respects, the $\eta$ coordinate
has been claimed to bring some advantages 
compared to $\sigma$ coordinates, especially
for the accuracy of the pressure-gradient force
computation, and for the assimilation of 
high-level data (Simmons and Burridge, 1981).
Hence, according to the warning signal obtained here,
the question of the relevance 
of the hybrid coordinate should 
legitimately have to be considered when 
implementing a real-case application
in EE system with $\pi$-type coordinate,
but cannot be answered with a reasonable 
certainty at this stage.


\section{Comments}
\label{sec_comment}

The intrinsic sensivity of SI schemes stability to the choice
of prognostic variables which has been demonstrated in this paper,
is totally independent of the space discretization, 
since the analyses have been performed in the 
space-continuous context. 
As a consequence, any numerical model built with
an intrinsically unstable variable will exhibit
an unstable behaviour, regardless of the 
space-discretization and staggering choices 
that may be done. However, it has 
been shown that an intrisincally
stable variable for height-based coordinates can 
become intrinsically unstable for mass-based 
coordinates because of the flow-dependent vertical 
metrics of this type of coordinates. 
Hence, it appears that for designing a SI 
numerical model in EE system, the 
choice of the prognostic variables cannot be made independently
of the choice of the vertical coordinate, even at the
level of space-continuous equations.

One could wonder if this dependency is specific to
the EE system or if it was already present in HPE 
systems, which are cast in  $\pi$-type coordinates.
First it should be noted that HPE system does not
offer a large latitude for the choice of prognostic
variables: for the problems examined here, 
the prognostic variable for the vertically 
integrated continuity equation is the only 
relevant one: it can be set to be $\pi_s$ or
$q=\ln(\pi_s)$.
The above numerical method applied to
the HPE system demonstrates the absence of
sensitivity to the choice of these prognostic
variables (not shown).
For instance, the instabilities reported by Simmons and Temperton (1997), 
which are the consequence
of using a hybrid coordinate, develop almost identically
for $\pi_s$ and $q=\ln(\pi_s)$ variables.
The sensitivity of the stability to the choice 
of the set of prognostic variables 
discussed here is thus a specificity of 
fully elastic nonhydrostatic systems (anelastic
systems have not been examined).

\section{Conclusion}

The stability of constant-coefficients SI schemes for the system of Euler Equations
has been examined from an theoretical point
of view, in deliberately simplified contexts to allow
tractable analyses.
The salient result of the study is that the stability
can be dramatically affected by
changes in the set of prognostic variables which
are used to design the SI scheme, and this, independently
of any space discretization.
It appears that the choice of the two "nonhydrostatic"
variables has to be carefully checked if an optimal 
stability is desired, especially for mass-based
coordinates. Non-optimal choices generally result
in growth rates incompatible with a practical NWP
use, as experienced in BHBG95. 
This sensitivity of the SI scheme stability to the
choice of prognostic variables
was not present in the HPE system.

If the SI reference 
surface-pressure deviates from its actual counterpart
for mass-based coordinates, hybrid coordinates
are found more unstable than the 
corresponding pure terrain-following coordinates,
as reported previously for HPE system, but the
practical consequences of this slight instability
need to be evaluated in more realistic contexts.

The analyses and results reported in this paper 
apply to constant-coefficients SI schemes, however, 
since only thermal and baric NL residuals are 
involved, the results extend identically to those
SI schemes belonging to the class (ii) of the 
introduction for which the reference 
temperature and pressure are horizontally
homogeneous (e.g. Thomas et al., 1998; Qian et al, 1998).

The practical interest of this theoretical study
could be considered as quite limited, since only a 
very small part of the NL terms due to the discrepancy
between the actual state ${\cal X}$ and the SI reference state 
$\cxst$ has been considered.
In this respect, the developments presented here
should rather be viewed as an endeavour to stress
that a special care must be exerted 
in the choice of prognostic variables for 
building a constant-coefficients SI EE system,
whatever type of coordinate is used.

In a more general way, there is another practical 
interest to this study: as a predictive tool, it
can serve for the validation of practical numerical
applications: in a similar way that analytically predicted 
stationnary orographic flows are commonly used for validating the
relevance of space-discretisation schemes, the analytically predicted 
stability can be used for a careful validation of
time-discretisation schemes. Moreover, if a "numerical-analysis" tool
is built, a complete set of cross-validations becomes possible
between analyses predictions, numerical-analyses diagnostics,
and observed numerical-model behaviour. This may significantly
help to improve the detection of anomalous behaviour in a
numerical model, or the prediction of the behaviour 
of alternative time discretizations (off-centered SI scheme, 
iterative treatments, etc...).

Nevertheless, as stated above, this study unavoidably leads 
to an overestimation of the stability in comparison 
to the likely stability for real-case conditions because
it retains only a very small part of all NL terms
arising from the discrepancy $({\cal M} - \clst)$. 
Among the most important sources of NL residuals, 
the spatial variability of the orography 
has been neglected so far.
Hence, there is no proof that the optimal 
set found here (e.g. $q$, $\pca$, $\nd$, 
and $\sigma$ coordinate) is still optimal,
or even reasonnably stable,
when a steep orography is introduced.
Further examination of this topic would
constitute an interesting extension for 
the present study.

\bigskip 
{\sl Acknowledgments}: Part of the research reported in this paper 
was supported by the ALATNET grant HPRN-CT-1999-00057 of the European 
Union TMR/IHP Programme. The authors would like to thank 
Dr. Claude Girard for fruitful discussions and for the 
experimental confirmation of the analytical results of Appendix B; 
and an anonymous reviewer for his helpful comments.

\newpage
\section*{Appendix A : List of Symbols}
               
\noindent vertical spatial operators in $\sigma$ coordinate:

   ${\cal G} X = \int_\sigma^1{(X/\sigma')} d\sigma'$ 

   ${\cal S} X = (1/\sigma) \int_0^\sigma{X} d\sigma'$ 

   ${\cal N} X = \int_0^1{X} d\sigma $

   ${\cal I} X = X$

   $\drti X = \sigma ( \dr / \dr \sigma) X$

\noindent miscellaneous symbols:

   $\bnabla$ : Horizontal gradient operator along constant surfaces of
   the considered vertical coordinate.
   
   ${\bf V}$ : Horizontal wind vector in 3D framework.
   
   $\nabla$ : ($\partial / \partial x$) along constant levels of
   the considered vertical coordinate in the 2D verical plane domain.
   
   $(\dr / \dr t)$ : Eulerian time-derivative
   
   $(d / dt)$ : Lagrangian time-derivative 

   $D$ : horizontal wind divergence ($\nabla .u$)
   
   $g$ : gravitational acceleration
   
   $p$ : true pressure

   $\pi$ : hydrostatic pressure
   
   $p_{00}$ : absolute reference pressure (1013.25 hPa)

   $R$, $C_p$, $C_v$ : dry air thermodynamic constants

   $u$ : horizontal wind component in the 2D framework
   
\newpage

\section*{Appendix B : Stability analysis of the EE in regular Gal-Chen coordinate}

The formalism for the derivation of the constant-coefficients 
SI scheme analysed here basically 
follows Caya and Laprise (1999). Notations are standard ones 
and taken identical to this paper unless specified. 
The unstretched Gal-Chen coordinate is defined by:

\vspace{-0.4in} 

\begin{equation}
\zeta = \frac{z - h_0}{\Ht - h_0}\Ht
\end{equation}

\noindent where $\Ht$ is the height of the domain, and $h_0$ is the 
height of the terrain. In the absence of orography ($h_0 \equiv 0$) 
we thus have:

\vspace{-0.4in} 

\begin{equation}
\zeta = z
\end{equation}

The general framework is the same as in the
above analyses, as depicted in section \ref{sec_fram},
and the thermal discrepancy between $\cxba$ and $\cxst$ states
is still noted $\alpha = (\tba - T^*)/T^*$. 
However, unlike for mass-based coordinates, the pressure variable
$q=\ln(p/p_{00})$ (not to be confused
with $q= \ln(\pi_s)$ for mass-coordinates systems in the main part of the paper)
needs to be defined in the whole space for both
$\cxba$ and $\cxst$ states. The hydrostatic equilibrium and stationnarity
of these states implies:

\vspace{-0.4in} 

\begin{equation}
\tba \frac{d \qba}{d z} = T^* \frac{d q^*}{d z}  = - \frac{g}{R}
\end{equation}

The SI time-discretized system is given by  
Eq. (46) -- (50) of Caya and Laprise, 1999:

\vspace{-0.4in} 

\begin{eqnarray}
\frac{d u}{d t} + R T^* \overline{\nabla q'}^t & = & - R T' \nabla q'  \\
\frac{d w}{d t} + R T^* \overline{\frac{ \dr q'}{\dr z}}^t
     - \frac{g}{T^*} \overline{T'}^t
& = & 
   - R T' \frac{ \dr q'}{\dr z} \\
\frac{d T'}{d t} - \frac{R T^*}{C_p} \frac{d q'}{d t} 
 + \frac{g}{C_p} \overline{w}^t
  & = & - \frac{R}{C_v} T' ( \nabla u  + \frac{\dr w}{\dr z}   ) \\
\frac{C_v}{C_p} \lc \frac{d q'}{d t} - \frac{g}{R T^*} \overline{w}^t\rc  
  + \lp \overline{\nabla u}^t +  \overline{\frac{\dr w}{\dr z}}^t \rp 
& = &   0 
\end{eqnarray}

\noindent where $T'=T-T^*$ and $q'=q-q^*$.  This system is exact in the 
sense that no term has been neglected so far. As can be
seen from the notation $\overline{ \rule[0.2cm]{3mm}{0mm}}^t$, 
the left part is treated in a centred implicit way
while the right part is treated explicitly.
For the analysis, a small pertubation ($\widetilde{T}$, $\widetilde{q}$)
around the actual state ($\overline{T}$, $\overline{q}$) must be introduced
and the complete system must be linearized around ($\overline{T}$, $\overline{q}$).
Equating the total variables $T$ and $q$ yields:

\vspace{-0.4in} 

\begin{eqnarray}
\widetilde{T} & = &  T' - \alpha T^*  \\
\widetilde{q} & = & q' - \frac{\alpha}{(1 + \alpha)} \frac{g}{R T^*} z
\end{eqnarray}

The terms which are non-linear in ($u$, $w$, $\widetilde{T}$, $\widetilde{q}$) 
are then dropped in the above system ($T'$ and $q'$ cannot be considered as small however).
The actual state $\cxba$ being a resting state, time derivatives are 
replaced by $(\dr / \dr t)$. However, a special care
must be taken to express $(d q' / dt)$ in term of $(d \widetilde{q} / dt)$, because
the vertical transport of $(\overline{q} - q^*)$ is a linear term which must
be retained. If an Eulerian scheme for the vertical advection is used, then:

\vspace{-0.4in} 

\begin{equation}
\frac{\dr q'}{\dr t} = \frac{\dr \widetilde{q} }{\dr t} + \frac{\alpha}{1+\alpha} \frac{g}{R T^*} w
\label{eq_dotq}
\end{equation}

\noindent where the advection term is treated explicitly. 
For a semi-Lagrangian vertical transport scheme,
the assumptions of perfect solution for the displacement equation, and of 
perfect interpolators, consistent with the space-continuous context, lead 
after some manipulation based on Taylor series expansions in space, 
to the same result than for the Eulerian advection scheme
as written in (\ref{eq_dotq}). 

\noindent The linearization of the above system around the $\cxba$ state
leads, after some algebraic manipulation to a form compatible with the 
analysis proposed in B03. The $\clba$ system then writes:

\vspace{-0.4in} 

\begin{eqnarray}
\frac{\dr u}{\dr t} & = & - R \tba \nabla \widetilde{q}  \\
\frac{\dr w}{\dr t} & = &   \frac{g}{\tba} \widetilde{T} 
- R \tba \frac{ \dr \widetilde{q} }{\dr z}\\
\frac{\dr \widetilde{q}}{\dr t} & = & \frac{g}{R \tba} w - 
\frac{C_p}{C_v} \lp \nabla u +  \frac{\dr w}{\dr z} \rp \\
\frac{\dr \widetilde{T}}{\dr t} & = &  
- \frac{R \tba}{C_v} \lp \nabla u +  
     \frac{\dr w}{\dr z} \rp.
\end{eqnarray}

\noindent The SI reference system $\clst$ is shown to be formally
identical, but with $\tba$ replaced by $T^*$.

\noindent We have $P=4$ and the normal modes 
of the time-continuous system are given  by:

\vspace{-0.4in} 

\begin{equation}
{\cal X}_j (x,z)  =  \cxha_j \; \exp(ikx)\, \exp(n z)   \;\; j \in (1, \ldots, 4)
\end{equation} 

\noindent where $n=(i \nu + 1/2) / \hba$ and  $\hba= R \tba/g$. 
It should be noted that the time-continuous normal modes of 
the $\clst$ system have a different vertical structure
$\exp(n^* z)$ with $n^*=(i \nu + 1/2) g/(R T^*)$. 
This discrepancy between the height-scales for 
the vertical growth of $\clba$ and $\clst$ 
time-continuous normal modes has some important
consequences, as discussed in section \ref{sec_pract}.

The analysis then proceeds as for mass-based coordinates by 
defining a numerical growth rate $\lambda$. 
For a given geometry $(k, \nu)$, the stability equation 
finally writes:

\vspace{-0.4in}
 
\begin{equation}
\frac{(\xxm)^4}{\Delta t^4} + \frac{(\xxm)^2}{\Delta t^2} c^2 \lc k^2 (\xxp) \xxa + n \nba (\xxp) \xxa 
+ n \xxa \frac{\alpha}{\hba} (\xxp - \lambda)\rc  +  k^2 N^2 c^2 \,\xxa^2 \xxb^2 = 0
\label{eq_GRT_MC2}
\end{equation}

\noindent where $\xxm$, $\xxp$, $\xxa$, $\xxb$, $c$, $H$, $N$ are defined as 
in (\ref{eq_def_xxp})--(\ref{eq_def_N}).

The asymptotic growth rate is then given by the same
equation (\ref{eq_AGR2}) as for mass-based coordinates, hence the 
conditions for unconditional stability are the same as well.
However, a major difference between Eqs. (\ref{eq_GRT_MC2}) 
and (\ref{eq_stru3}) is the presence of 
the last complex term in the bracket of (\ref{eq_GRT_MC2}). 
The solution of
this equation for finite time-steps shows that this 
term, being complex,  is responsible of an instability when $\alpha \neq 0$.
The instability is generally small, but
is found to become quite significant when $k \hba \approx \nu \approx 1$,
but, as stated above, it always disappears for long time-steps.


\newpage

\section*{References}

\begin{description}

\item Bénard, P., 2003:
      Stability of Semi-Implicit and Iterative Centered-Implicit Time
      Discretizations for Various Equation Systems Used in NWP.
      {\em Mon. Wea. Rev.}, {\bf 131}, 2479-2491.
      
\item Bubnov\'a, R., G. Hello, P. B\'enard, and J.F. Geleyn, 1995:
      Integration of the fully elastic equations cast in the hydrostatic
      pressure terrain-following coordinate in the framework of the
      ARPEGE/Aladin NWP system.
      {\em Mon. Wea. Rev.}, {\bf 123}, 515-535.
      
\item Caya, D., and R. Laprise, 1999:
      A semi-implicit semi-Lagrangian regional climate model: the Canadian RCM.
      {\em Mon. Wea. Rev.}, {\bf 127}, 341-362.
      
\item C\^{o}t\'{e}, J., M. B\'eland, and A. Staniforth, 1983:
      Stability of vertical discretization schemes for semi-implicit
      primitive equation models: theory and application.
      {\em Mon. Wea. Rev.}, {\bf 111}, 1189-1207.
      
\item C\^{o}t\'{e}, J., S. Gravel, and A. Staniforth, 1995:
      A generalised family of schemes that eliminate the spurious resonant
      response of semi-Lagrangian schemes to orographic forcing.
      {\em Mon. Wea. Rev.}, {\bf 123}, 3605-3613.
\item Eckart, C., 1960:
      {\em Hydrodynamics of oceans and atmospheres}, Pergamon, 290 pp.
\item Gal-Chen, T, and R. Somerville, 1975:
      On the use of a coordinate transformation for the solution 
      of the Navier-Stokes equations variable.
      {\em J. Comput. Phys.}, {\bf 17(2)}, 209-228.
\item Laprise, R., 1992:
      The Euler equations of motion with hydrostatic pressure as an
      independent variable.
      {\em Mon. Wea. Rev.}, {\bf 120}, 197-207.
\item Laprise, R., D. Caya, G. Bergeron, M. Gigu\`ere 1995:
      The formulation of the Andr\'e Robert MC2 
      (mesoscale compressible community) model.
      {\em Atm. Ocean, special issue: the Andr\'e J. Robert memorial volume }, 195-220.
\item Qian, J.-H., F. H. M. Semazzi, and J. S. Scroggs, 1998:
      A global nonhydrostatic semi-Lagrangian atmospheric model with
      orography.
      {\em Mon. Wea. Rev.}, {\bf 126}, 747-771.
\item Robert, A. J., J. Henderson, and C. Turnbull, 1972: An implicit time
      integration scheme for baroclinic models of the atmosphere . 
      {\em Mon. Wea. Rev.}, {\bf 100}, 329-335.
\item Simmons, A. J., and D. Burridge, 1981:
      An Energy and Angular-Momentum Conserving Vertical Finite-Difference Scheme
      and Hybrid Vertical Coordinates.
      {\em Mon. Wea. Rev.}, {\bf 109}, 758-766.      
\item Simmons, A. J., C. Temperton, 1997:
      Stability of a two-time-level semi-implicit integration scheme
      for gravity wave motion.
      {\em Mon. Wea. Rev.}, {\bf 125}, 600-615.      
\item Simmons, A. J., B. Hoskins, and D. Burridge, 1978:
      Stability of the semi-implicit method of time integration.
      {\em Mon. Wea. Rev.}, {\bf 106}, 405-412.
\item Semazzi, F. H. M., J. H. Qian, and J. S. Scroggs, 1995:
      A global nonhydrostatic semi-Lagrangian atmospheric model without
      orography.
      {\em Mon. Wea. Rev.}, {\bf 123}, 2534-2550.
\item Tanguay M., A. Robert, and R. Laprise, 1990:
      A semi-implicit semi-Lagrangian fully compressible regional forecast model.
      {\em Mon. Wea. Rev.}, {\bf 118}, 1970-1980.
\item Thomas, S.J., C. Girard, R. Benoit, M. Desgagn{\'e}, and P. Pellerin, 1998:
      A new adiabatic kernel for the MC2 model.
      {\em Atmos. Ocean}, {\bf 36 (3)}, 241-270.
\end{description}

\newpage

\section*{List of Figures}

 Fig. 1:  Analytic growth rates (GR) 
 for the simplified problem as a function of the 
 nonlinearity parameter $\alpha$.
  Thick line: asymptotic GR for variable $\nd$, 
  solid line: asymptotic GR for variable $\hd$,
  dotted line: practical GR for variable $\nd$.
  dashed line: practical GR for variable $\hd$
  (see text for the meaning of practical GR).

\medskip

Fig. 2: Numerical growth rate  
 (left axis) for the realistic thermal profile 
 (dotted line - pressure on right axis),as a function of $T^*$. 
 From thin to thick lines: 10, 20, 30 levels.
  solid line: variable $\hd$,
  dashed line: variable $\nd$.
  
\medskip

Fig. 3:  Numerical growth rate  
 (left axis) for the realistic thermal profile 
 (dotted line - pressure on right axis),as a function of $T^*$. 
  Thick lines: $\pi_s^*$ = 1213.25 hPa,
  thin lines : $\pi_s^*$ = 813.25 hPa,
  solid line: variable $\hpca$ with $\sigma$ coordinate,
  dashed line: variable $\pca$ with $\eta$ coordinate.

\newpage

\addtocontents{lof}{\protect\vspace{0.5cm}}

\addtocontents{lof}{\protect\vspace{0.5cm}}

\addtocontents{lof}{\protect\vspace{0.5cm}}


\end{document}